\documentclass[aps,amsfonts,pre,preprint,superscriptaddress,showpacs]{revtex4-1}

\usepackage[dvips]{graphicx}
\usepackage{amssymb,amsfonts,amsmath,bm}
\usepackage[usenames]{color}
\usepackage{bm}

\newcommand{\remove}[1]{}

\begin{document}

\title{Localizing softness and stress along loops in three-dimensional topological metamaterials}

\author{Guido Baardink}
\affiliation{Instituut-Lorentz, Universiteit Leiden, 2300 RA Leiden, The Netherlands}
\author{Anton Souslov}
\affiliation{Instituut-Lorentz, Universiteit Leiden, 2300 RA Leiden, The Netherlands}
\author{Jayson Paulose}
\affiliation{Instituut-Lorentz, Universiteit Leiden, 2300 RA Leiden, The Netherlands}
\affiliation{Departments of Physics and Integrative Biology, University of California, Berkeley, CA 94720, USA }
\author{Vincenzo Vitelli}
\affiliation{Instituut-Lorentz, Universiteit Leiden, 2300 RA Leiden, The Netherlands}

\date{\today}

\begin{abstract}
Topological states can be used to control the mechanical properties of a material
along an edge or around a localized defect. The surface rigidity of elastic
networks is characterized by a bulk topological invariant called the
polarization; materials with a well-defined uniform polarization display a
dramatic range of edge softnesses depending on the orientation of the polarization relative to the
terminating surface. 
However, in all three-dimensional mechanical metamaterials proposed to date, the topological edge modes are mixed with bulk soft modes and so-called Weyl loops. Here, we report the design of a gapped
3D topological metamaterial with a uniform polarization that displays a corresponding asymmetry
between the number of soft modes on opposing surfaces and, in addition, no bulk
soft modes. We then use this construction to localize topological soft modes in interior
regions of the material by including defect structures---dislocation
loops---that are unique to three dimensions. We derive a general formula that
relates the difference in the number of soft modes and states of self-stress localized along the dislocation loop to the handedness of the 
vector triad formed by the lattice polarization, Burgers vector, and dislocation-line direction. Our
findings suggest a novel strategy for pre-programming failure and softness localized
along lines in 3D, while avoiding extended periodic failure modes
associated with Weyl loops.
\end{abstract}

\maketitle

\section*{Introduction}
Mechanical metamaterials can control softness via 
a balance between the number of degrees of freedom of their components or nodes
and the number of constraints due to connections or links~\cite{Bertoldi2010,Sun2012,Kane2014,Chen2014,Lubensky2015, Meeussen2017, Rocks2017, Yan2017, Coulais2017}.
This balance, first noted by Maxwell~\cite{Maxwell1864} and later explored by Calladine~\cite{Calladine1978}, is termed \emph{isostaticity}.
In isostatic materials, softness can manifest itself via large-scale deformations, for example
as uniform Guest-Hutchinson modes~\cite{Guest2003, Kapko2009} or via periodic soft deformations, corresponding to so-called Weyl modes~\cite{Rocklin2016, Stenull2016}.
Uniform softness can be exploited to create extraordinary mechanical response~\cite{Florijn2014}, such as materials with a negative Poisson's ratio~\cite{Lakes1987,Bertoldi2010,Sun2012}.
Alternatively, localized softness has been programmed into isostatic materials in one and two dimensions
via a topological invariant called the polarization~\cite{Kane2014,Lubensky2015} that controls mechanical response and stress localization~\cite{Rocklin2016a} at an edge [including the edge of a disordered sample~\cite{Sussman2016}], an interface, or bound to a moving soliton~\cite{Chen2014}.
These mechanical~\cite{Nash2015, Susstrunk2015, Kariyado2015, Mousavi2015, Khanikaev2015,Yang2015,Huber2016} examples of topological metamaterials~\cite{Rechtsman2013, Lu2014, Wang2015} exhibit a general feature of topological matter~\cite{Hasan2010, Bernevig2013}: a correspondence between integer invariants in the bulk and response at a boundary.
Large-scale and localized deformations are deeply intertwined, as can be seen in demonstrations in which topological edge softness is created or destroyed by applying large uniform strains~\cite{Rocklin2017}.
A combination of topological polarization and localized defects can be used to program in softness or failure at a specified region in the material~\cite{Paulose2015,Paulose2015a}.

Although there are a number of examples of isostatic periodic structures in one, two, and three dimensions, the three-dimensional case is unique because all prior realizations of three-dimensional isostatic lattices include large-scale periodic deformations along continuous lines in momentum space~\cite{Paulose2015a, Stenull2016, Bilal2017}.
These Weyl lines define families of periodic soft modes in the material bulk
and contain a number of modes, which scales with the linear size of the structure.
As Ref.~\cite{Bilal2017} explores, Weyl lines can be useful to create a metamaterial surface with anisotropic elasticity, but in order to create a material whose top surface is much softer than the bottom, it proves necessary to collapse two Weyl lines on top of each other. 
An alternative would be to find a metamaterial without Weyl lines.
However, these Weyl lines are generic and have a topological character which ensures that they cannot be annihilated locally---a single Weyl loop can only be destroyed by shrinking it to a point.
This presents a challenge in three-dimensional isostatic metamaterial design: 
to achieve a gapped topological material analogous to those in two dimensions~\cite{Kane2014,Lubensky2015, Rocklin2017}, in which softness can be controlled and localized.

In this work, we design gapped topological materials by exploring the parameter space of the generalized stacked kagome lattice. 
The mechanical response of this gapped structure is characterized by a nonzero topological polarization oriented along the $z$-axis. 
This topological polarization $\bm P$ can be exploited 
to localize soft modes in the material bulk by introducing topological defects within the lattice structure called \emph{dislocation loops}.
These dislocations are characterized by a topological invariant called the Burgers vector $\bm b$.
Along the dislocation, we show that the topological charge characterizing the softness or rigidity of the lattice (with unit cell volume $V_{cell}$) depends on the orientation $\bm{\hat\ell}$ of the dislocation line and is given by $\bm P\cdot(\bm b\times\bm{\hat\ell})/V_{cell}$ per unit length.

\begin{figure}[th!]
\centerline{\includegraphics{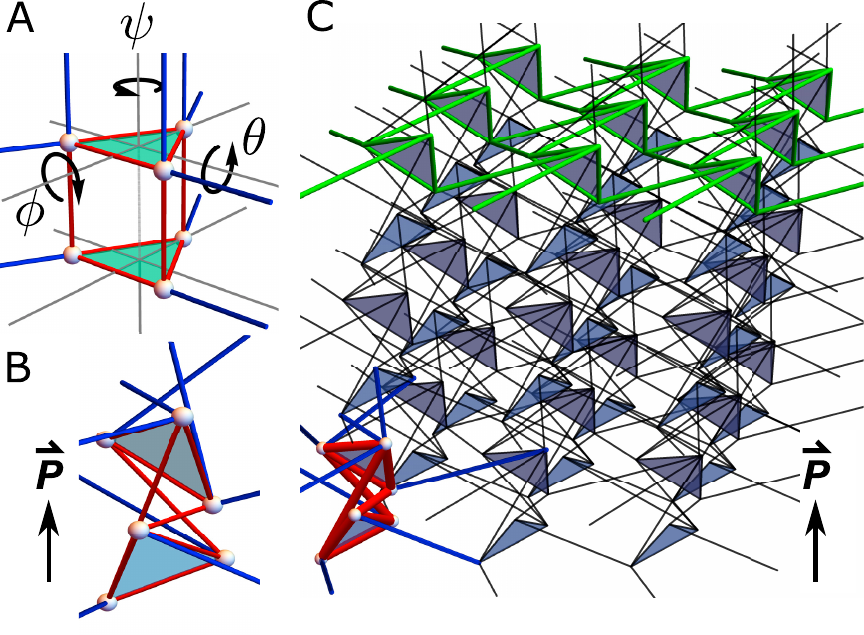}}
\caption{\label{Fig1} Architecture of a gapped 3D stacked kagome lattice.
(A) We consider the doubled unit cell of a vertically stacked kagome lattice and rotate
the top constituent triangle around three perpendicular axes by angles $(\phi, \theta, \psi)$ [the bottom triangle is rotated by $(- \phi, - \theta, - \psi)$].
(B) For the choice $(\phi, \theta, \psi) = (\pi/3, 0, \pi/3)$ of these parameters, we obtain a unit cell that corresponds to a (gapped) 3D lattice with a topological polarization $\bm{P}$ and without Weyl lines. 
(C) Architecture of the gapped metamaterial with one unit cell emphasized in the lower left and a 2D kagome sub-lattice surface highlighted in green on the top. }
\end{figure}

\section*{Gapping the stacked kagome lattice}
We examine the mechanics of metamaterial structures by using a lattice model for the displacements of nodes and strains of the links. Within this set-up, softness corresponds to displacements with small strains, and rigidity to strains that induce only small displacements.
We place a particle (or ball) at each vertex and connect the neighboring particles by linear springs. Such models capture the small-strain response of realistic structures that are either 3D-printed from soft polymers~\cite{Bilal2017, Paulose2015a} or assembled from construction sets~\cite{Chen2014} or laser-cut components~\cite{Paulose2015}. 

The mechanics of this ball-spring model are captured via the linearized equation of motion 
$\bm{\ddot X} = - D \bm{X}$, where $\bm{X}\equiv(\bm{x}_1,\dots,\bm{x}_N)$ is a $d \times N$-dimensional vector containing the displacements of all $N$ particles in $d$-dimensions relative to their equilibrium positions.
For a given lattice geometry, we calculate the dynamical matrix $D$, which relates the forces exerted by springs to displacements of the particles. For simplicity, we work in units in which all particle masses and spring constants are one.
In the linear regime, we can find the dynamical matrix by first relating the $N_B$-dimensional vector of spring extensions $\bm{S}\equiv(s_1,\dots,s_{N_B})$ to the displacements via $\bm{S}=R\bm{X}$.
The rigidity matrix $R$ contains $dN\times N_B$ entries determined by the equilibrium positions of the particles and the connectivity of the lattice.
In combination with Hooke's law, the matrix $R$ lets us calculate the dynamical matrix $D$ via the relation 
$D = R^\mathsf{T}R $~\cite{Calladine1978,Sun2012}. Although the rigidity theory in terms of the matrix $R$ offers an equivalent description to the dynamical matrix, it contains additional physical insight.

The null spaces (kernels) of the rigidity matrix and its transpose give us information about lattice mechanics. From the relation $\bm{S}=R\bm{X}$, we note that $\ker R$ contains \emph{soft modes}, i.e., collective displacements that (to lowest order) do not stretch or compress any of the springs. On the other hand, the matrix $R^\mathsf{T}$ relates forces on particles $\bm{\ddot X}$ to spring strains via $\bm{\ddot X} = - R^\mathsf{T}\bm{S}$. From this relation, one notes that $\ker R^\mathsf{T}$ contains combinations of spring tensions that do not give rise to particle forces. These configurations are dubbed \emph{states of self-stress} because they define load-bearing states, which put the lattice under a static tension~\cite{Calladine1978}. Localized states of self-stress can concentrate uniformly-applied external loads, which can lead to selective buckling in pre-programmed regions~\cite{Paulose2015a}.

To calculate soft modes (states of self-stress), we need to numerically solve the configuration-dependent equation $R\bm{X} = \bm{0}$ ($R^\mathsf{T}\bm{S} = \bm{0}$). 
However, a mathematical result called the rank-nullity theorem lets us compute the difference between the number of zero modes $N_0 \equiv \mathrm{null} R$ and the number of states of self-stress $N_B \equiv \mathrm{null} R^\mathsf{T}$~\cite{Calladine1978,Sun2012,Lubensky2015}. This difference, the rigidity charge $\nu \equiv N_0 - N_B$, is given by the dimensionality of the $R$ matrix: $\nu = dN - N_B$,
and can only change if balls or springs are either added or removed.
We focus on the special case of isostatic, or Maxwell, lattices defined by $\nu = 0$ when the system is considered under periodic boundary conditions. These lattices are marginally rigid and exhibit a symmetry between zero modes and states of self-stress.

To design metamaterials based on simple, repeated patterns, we focus on periodic structures. Periodicity allows us to explicitly calculate zero modes and states of self-stress in a large sample. We begin with a highly symmetric, ``undeformed'' lattice and explore its configuration space by changing the positions (but not the connectivity) of the nodes.
The specific geometry that we consider is illustrated in Fig.~1: this stacked kagome lattice has a coordination number $z\equiv 2N_B/N = 2d= 6$ and $\nu = 0$: the lattice is isostatic.
This lattice is based on the two-triangle unit cell shown in Fig.~\ref{Fig1}A, which corresponds to two unit cells of the kagome lattice stacked on top of each other. We deform the lattice via the orientations of the triangles, which are governed by the three angles $(\phi,\theta,\psi)$ of rotation around the $(x,y,z)$-axes, respectively. 
Within this structure, we fix the connectivity and explore a range for each of the three angles from $-\pi/2$ to $\pi/2$. In Fig.~\ref{Fig1}B, we show the deformed unit cell for a particular structure in this region, which corresponds to the choice $(\phi,\theta,\psi) = (\pi/3,0,\pi/3)$. This unit cell builds the periodic geometry in Fig.~\ref{Fig1}C. We proceed to quantitatively demonstrate that this metamaterial is gapped.

We calculate the spectrum of normal modes by considering plane-wave solutions of the form $x_{\alpha,\bm{\ell}}=e^{i\bm \ell\cdot\bm{k}}x_{\alpha}$, where the index $\alpha$ refers to a particle within one unit cell and the lattice index $\bm\ell$ enumerates different unit cells within a lattice.
The three-component wavevector $\bm{k}$ is periodic in each component with $-\pi\leq k_x, k_y,k_z\leq\pi$. This collection of points forms the first Brillouin zone of the lattice. 
The Bloch representation of the rigidity matrix in this plane-wave basis is $R_{\alpha\beta}(\bm{k})=R_{\alpha\beta}e^{i(\bm\ell_\alpha-\bm\ell_\beta)\cdot\bm{k}}$. A zero mode at wave-number $\bm{k}$ is a vector of unit-cell displacements $x_\alpha$ that solves $R_{\beta\alpha}(\bm{k})x_\alpha=0$. Thus, within this setting, zero modes correspond to the zeros of the complex function $\det R(\bm{k})$. By examining this function, we find the zero modes for different configurations of the lattice.

\begin{figure*}[th!]
\centerline{\includegraphics{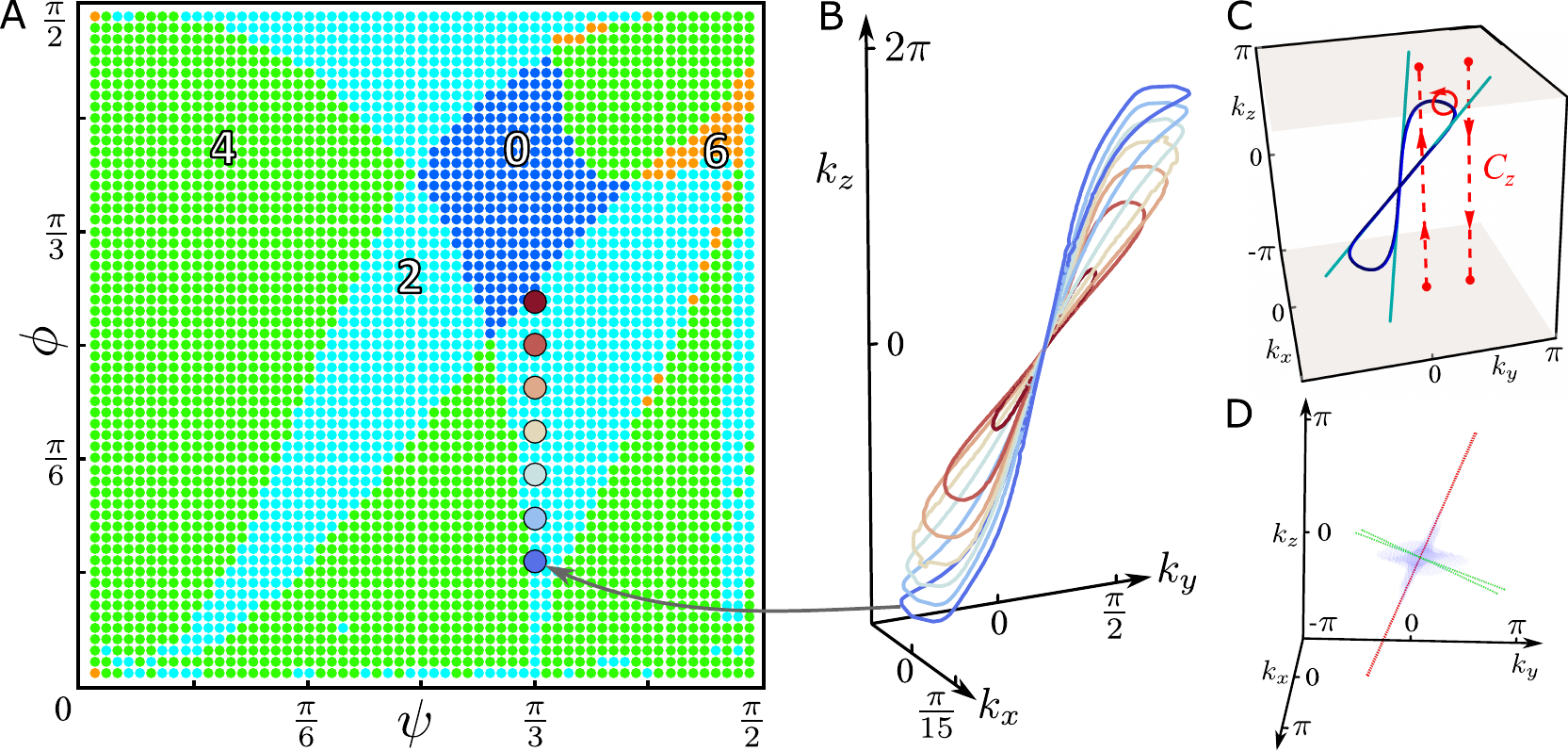}}
\caption{\label{Fig2} Phase diagram of stacked kagome lattices.
(A) The number of soft directions in $\phi$-$\psi$ parameter space for $\theta = 0$ (see Fig.~\ref{Fig1}A).
The region in dark blue corresponds to a gapped topological lattice, whereas the rest corresponds to lattices containing two (cyan), four (green) or six (orange) Weyl loops. 
(B) Weyl loops along the line $(\theta,\psi)=(0,\pi/3)$, corresponding to magnified points in (A). As the gapped region is approached we observe Weyl loops shrinking towards the origin until they disappear.
(C) An example of the soft-mode structure in reciprocal space for a lattice with a single two-Weyl-loop structure (in blue), which corresponds to the cyan point $(\phi,\theta,\psi)=(5 \pi/12,0,\pi/4)$ in (A). 
The integration of a phase around any contour (e.g., solid red circle) enclosing the Weyl loop 
gives an integer winding number that we use to check the topological nature of the Weyl loop.
Note that this contour can be deformed into a pair of contours ($C_z$, dashed red lines) traversed in opposite directions: this illustrates how the winding number across the Brillouin zone changes as the Weyl loop is traversed.
Near the origin $(k_x,k_y,k_z) = (0,0,0)$, the loop manifests itself via soft directions (teal).
(D) The Brillouin zone of a lattice corresponding to the blue point $(\phi,\theta,\psi)=(\pi/3,0,\pi/3)$ in (A). Far away from the origin, the lattice is checked to be gapped. At the origin, the softest regions (blue region) align themselves with the soft directions of lattices with $(\theta,\psi)=(0,\pi/3)$ (red) or $(\phi,\theta)=(\pi/3,0)$ (green).}
\end{figure*}

\clearpage

We numerically evaluate zero modes for the stacked kagome lattice (see Supporting Information [SI] for details) and for most values of the angles $(\phi,\theta,\psi)$ find collections of zero modes along compact loops in $\bm{k}$-space [Fig.~\ref{Fig2}A]. These Weyl loops, shown in Fig.~\ref{Fig2}B--C, 
are analogous to one-dimensional nodal lines~\cite{Volovik2007, Burkov2011, Lu2013} and experimentally observed zero-dimensional Weyl points~\cite{Xu2015,Lu2015} in three-dimensional electronic semimetals and photonic crystals.
In isostatic lattices, the number of Weyl loops is always even, because the materials' time-reversal symmetry maintains $\bm{k}\to-\bm{k}$ reflection symmetry in the Brillouin zone---each Weyl loop comes in a pair with its reflected partner~\cite{Stenull2016}. 
Furthermore, these loops attach to the origin of the Brillouin zone along soft directions, i.e., the lines tangent to the Weyl loops at the origin. We count the number of loops by looking at soft directions in the neighborhood of the origin and, in Fig.~\ref{Fig2}A, plot this count in a slice of parameter space. In this phase diagram, we note regions with up to six different loops. 
Strikingly, the middle of the diagram displays a region in which no Weyl loops exist [Fig.~\ref{Fig2}A,D].

We conclude that although Weyl loops are generic, the stacked kagome lattice also exhibits \emph{gapped} configurations which contain no Weyl loops. 
In lattices with Weyl loops, the number of soft modes scales as the linear size of the system, whereas gapped lattices have only the three uniform Guest-Hutchinson modes~\cite{Guest2003, Kapko2009}. 
We are not aware of any other realizations of a three-dimensional isostatic lattice which is gapped.
In the next section, we address the implications of the existence of a gap for the topological characterization of mechanical networks.

\section*{Topological rigidity in three dimensions}
For the gapped isostatic lattice, an integer topological invariant called the polarization can be computed from the bulk phonon spectrum~\cite{Kane2014}. This winding-number invariant is only well defined in the gapped case: in Weyl lattices, the closure of the gap prevents a consistent definition. That is, when Weyl loops are present, the polarization changes depending on the choice of contour in the Brillouin zone. The mechanical consequences of polarization are apparent by looking at the spectrum of the material's soft surface waves. The polarization controls which surfaces have more soft modes: this is the mechanical version of the bulk-boundary correspondence principle.

\begin{figure}[th!]
\centerline{\includegraphics{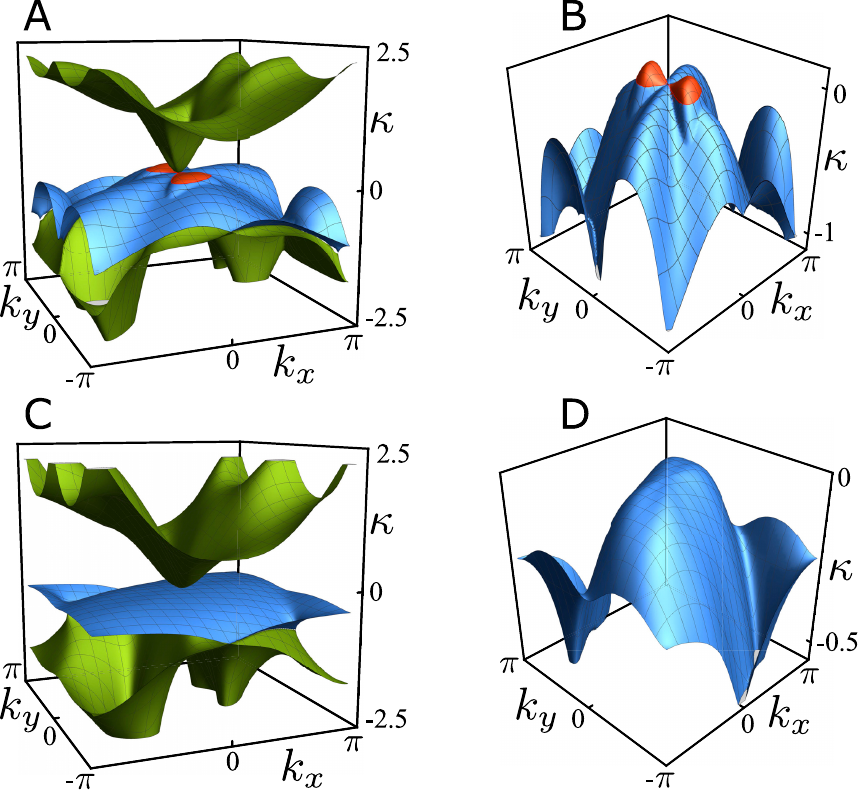}}
\caption{\label{Fig3}  Surface modes for a gapped polarized lattice (C--D) compared to
a lattice with Weyl lines (A--B).
A surface mode in a plane perpendicular to the $z$-axis (see, e.g., Fig.~\ref{Fig1}C) decays exponentially into the material bulk with 
amplitude $e^{- \kappa z/|\bm{a}_3|}$, where $\kappa$ is the inverse penetration depth.
Lattice with Weyl lines $(\phi,\theta,\psi)= (5 \pi/12, 0, \pi/4)$: (A) Inverse penetration depth for three soft surface modes over the entire surface Brillouin zone (sBZ).
For small wavevectors $(k_x, k_y)$, these modes correspond to the three acoustic phonons.
(B) Zoom on the middle mode in (A): the part of wavevector space for which
the mode lives on the bottom side is colored in red, whereas the part corresponding
to the top side is in blue. The projection of the Weyl line from the bulk 
onto the sBZ corresponds to a line with infinite penetration depth ($\kappa = 0$) and separates the red and blue regions.
Gapped polarized lattice $(\phi,\theta,\psi)= (\pi/3, 0, \pi/3)$: (C) Inverse penetration depth for the same three modes as in (A), but for a different choice of parameters,
corresponding to the blue region in Fig.~\ref{Fig2}A.
(D) Zoom-in on the middle mode in (C), which [in contrast to the mode in (B)] is localized on the top of the lattice only:
it is blue throughout. Significantly, the bulk ($\kappa = 0$) modes that
correspond to the Weyl loop are absent for this lattice.}
\end{figure}

\clearpage

For the gapped polarized lattice, a polarization vector $\bm P$ can be computed via $\bm P=\sum_i m_i\bm a_i - \bm{d}_0$, where $\bm{a}_i$ are the three lattice vectors 
and, in our case, the unit-cell dipole $\bm{d}_0 = -3 \bm a_3/2$. This dipole is computed via the expression $\bm{d}_0 \equiv - \sum_{b}\bm{r}_b$, where $\bm{r}_b$
are the positions of the bond centers relative to the center of mass of the unit cell~\cite{Kane2014}.
The three coefficients $m_i$ are winding numbers computed by integrating the phase change of the complex function $\det R(\bm k)$ across a straight-line contour crossing the Brillouin zone:
\begin{equation}
m_i \equiv - \frac{1}{2\pi i}\oint_{C_i}d k_i \frac{d}{d k_i}\ln\det R(\bm k), \label{eq:m}
\end{equation}
where $C_i$ is a closed contour in the $\bm{\hat k_i}$ direction~\cite{Kane2014}. This integral is well defined for contours along which $\det R(\bm k) \neq 0$. The integers $m_i$ depend smoothly on the choice of contour and are constant. In gapped lattices, the determinant is non-zero everywhere except for the origin $\bm k=0$. As a consequence, the winding numbers are independent of the chosen contour and the polarization is a topological invariant. On the other hand, Weyl loops partition the space of straight-line contours into lines going through the inside (outside) of the loop. Because $\det R(\bm k)=0$ along the loops, contours on either side of the loop can have different winding numbers $m_i$. Note that the combination of any two such contours taken in opposite directions can be smoothly deformed (without intersecting the Weyl line) into a small circle enclosing the Weyl line, as shown in Fig.~\ref{Fig2}C. The winding number $m_\mathrm{W}$ around this small circle is an invariant and equal to the difference between $m_i$ for $C_i$ on two sides of the loop. This topological protection guarantees that a single Weyl loop cannot be destroyed, which explains why Weyl loops only vanish by shrinking to the origin within the phase diagram in Fig.~\ref{Fig2}A. In summary, Weyl lines are protected by a winding number, and gapped lattices can have a well-defined topological polarization.

Bulk-boundary correspondence states that the topological invariants computed in the bulk can have significant effects on the mechanics of a sample with boundaries. We demonstrate this correspondence by computing the topological invariants and the spectrum of soft edge states in the stacked kagome lattice. 
In the bulk of the Weyl lattice corresponding to $(\phi,\theta,\psi)=(5\pi/12,0,\pi/4)$ (c.f., Fig.~\ref{Fig2}A,C), the winding number around the loop is $m_\mathrm{W}=- 1$. The $m_3$ for contours inside and outside of the loop differ by one. To see the boundary counter-part of the correspondence, we compute the spectrum of soft edge states for this lattice with a stress-free surface parallel to the $xy$-plane (see Fig.~\ref{Fig1}C). In Fig.~\ref{Fig3}A--B,
we show the (signed) inverse penetration depth $\kappa$ for soft surface modes in the two-dimensional Brillouin zone for wavevector $(k_x,k_y)$. These plane-wave solutions of $\det R = 0$ correspond to soft modes of the form $e^{i\bm\ell\cdot(\bm{k}+i\bm{\kappa})}\bm{x}_\alpha$. The middle mode of Fig.~\ref{Fig3}A (see zoom in Fig.~\ref{Fig3}B) changes sides: for $\kappa > 0$ (red region), the soft mode is attached to the bottom surface and decays upwards into the bulk, whereas for $\kappa < 0$ (blue region), the mode is localized on the top surface. 
For the line corresponding to the projection of the Weyl loop onto the two-dimensional Brillouin zone, the penetration depth is infinite and $\kappa = 0$, because the Weyl lines are soft modes in the bulk.
The difference in $m_3$ between inside and outside of the Weyl loops is  the bulk invariant $m_\mathrm{W} = - 1$, which corresponds to the difference between the number of soft surface modes across the projected Weyl loop. This connection follows from Cauchy's argument principle for Eq.~(\ref{eq:m}), which states that the third winding numbers $m_3$ count, up to a constant, the number of modes with zero energy ($\det R = 0$) at a boundary. These observations confirm bulk-boundary correspondence.

A similar correspondence exists in the gapped lattice. There, the invariant polarization pointing along the $z$-axis is given by $\bm P = \bm a_3/2$, which we computed for parameters $(\phi,\theta,\psi)=(\pi/3,0,\pi/3)$. 
To understand the effect of $\bm P$ on the boundary, note in analogy with electromagnetism that the rigidity charge $\nu^S$ in region $S$ is related to the flux of polarization $\bm P$ through the region's boundary $\partial S$. In a nearly-uniform, gapped lattice~\cite{Kane2014}
\begin{equation}
\nu^S{=}\oint_{\partial S}\frac{dA}{V_{cell}}\bm{\hat n}\cdot \bm P,
\label{eq:nu}
\end{equation}
where $\bm{\hat n}$ is the boundary's inward normal and $V_{cell}=\det(\bm a_1,\bm a_2,\bm a_3)$ is the volume of a unit cell. 
The difference between the rigidity charges at the top versus the bottom is determined by $\bm P$: for the stacked kagome, we expect one more band of soft modes along the top surface relative to the bottom. 
The total number of soft surface modes is three as a result of three bonds being cut per unit cell~\cite{Kane2014}.
We plot the inverse penetration depth $\kappa$ for this extra band in Fig.~\ref{Fig3}C--D within the two-dimensional Brillouin zone. 
Unlike the Weyl lattice, the whole middle band in Fig.~\ref{Fig3}C is blue: the top surface has more soft modes and is, therefore, softer.

Equation~(\ref{eq:nu}) shows that topological polarization acts analogously to an electric polarization. Inside a homogeneous polarized material, the charge is zero. However, if homogeneity is broken by the presence of boundary or \emph{defects}, charge can accumulate at these spots.

\begin{figure*}[h!]
\centerline{\includegraphics{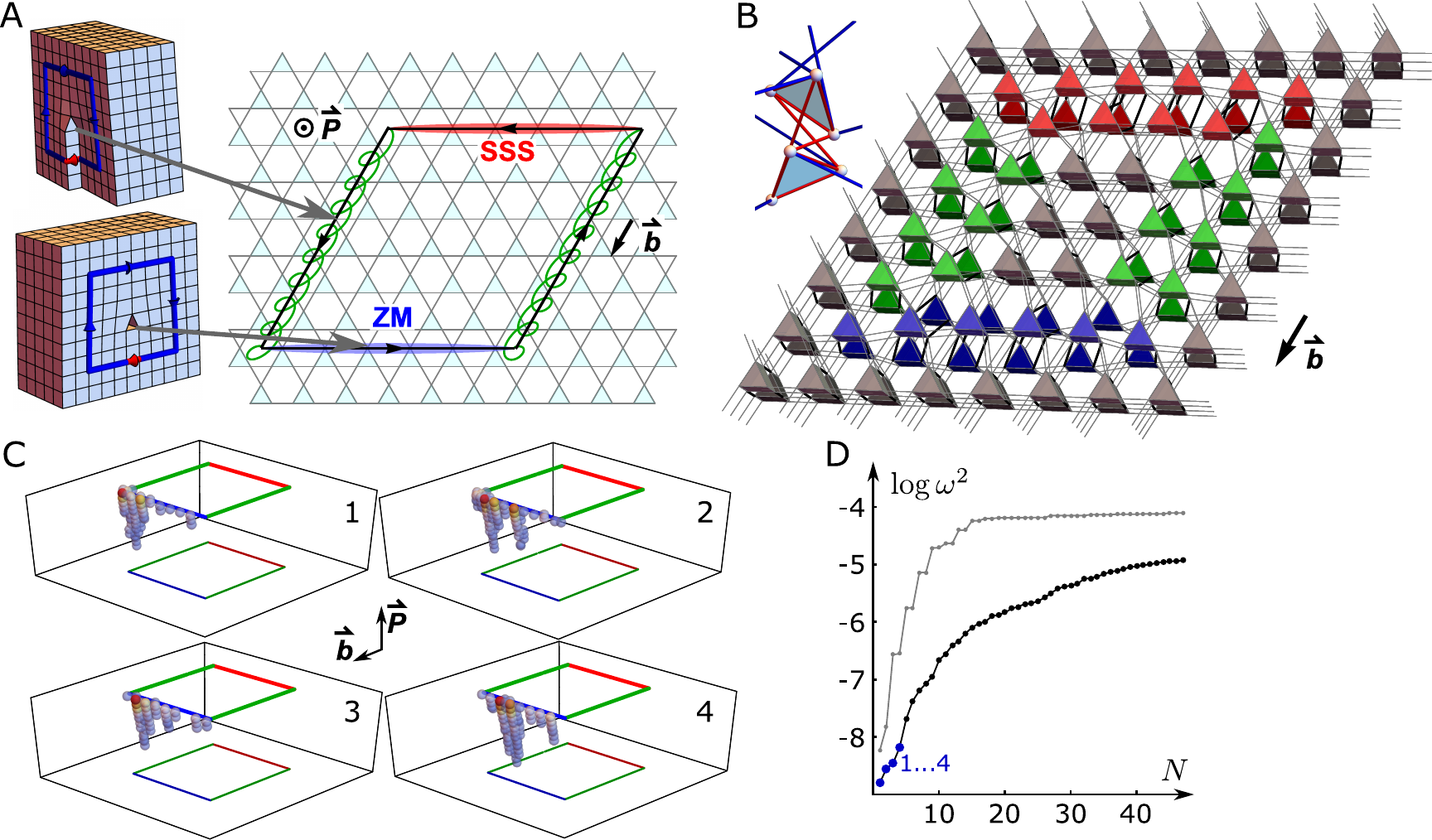}}
\caption{\label{Fig4} Dislocation loops can act as dipoles of topological charge.
(A)~Architectures of a dislocation loop in a periodic lattice:
in a screw dislocation, the burgers vector $\bm{b}$ is parallel to the dislocation line
whereas in an edge dislocation, the burgers vector and the dislocation line are perpendicular.
 A dislocation loop can combine edge and screw dislocations along its contour. In a gapped polarized lattice (with polarization $\bm{P}$), a dislocation line separates in space an edge dislocation segment that carries zero modes (ZM, blue) from an edge dislocation segment that carries states of self-stress (SSS, red), for example via screw dislocations (green) which carry no such charges. The net topological charge, defined as the difference between ZM and SSS, is zero when summed over a dislocation loop contour. Nevertheless, the dislocation loop carries a topological charge dipole, which is in this example is parallel to the Burgers vector $\bm{b}$.
(B)~Geometry of the dislocation loop. Each prism represents a unit cell with triangles oriented according to $(\phi,\theta,\psi)=(\pi/3,0,\pi/3)$ as shown in the inset. (C)~Numerical results for the four lowest-frequency phonons (excluding the trivial translations) in a large ($36\times36\times18$ unit cells) gapped lattice that has a dislocation loop: warmer color signifies a larger displacement within that unit cell. Plotted are only those unit cells that have displacements above a cutoff of 20\% of the maximum. 
Note the localization of the softest modes to the near side of the loop, in accordance with part A and the polarization $\bm{P}$.
(Caption continued on next page.)
}
\end{figure*}
\clearpage
\addtocounter{figure}{-1}
\begin{figure*}[th!]
\caption{(continued) 
(D)~The common logarithm of lowest mode frequencies $\log \omega^2$ for the $N$-th lowest mode, plotted versus $N$, comparing two large samples (same size as C).
The mode frequencies are significantly lower for the dislocated lattice (black), than in the non-dislocated case (gray).
Without a dislocation, the lowest eigenmodes are the extended plane-wave acoustic phonons, whereas
with the dislocation, these modes include both the acoustic phonons and the modes localized along the dislocation loop (in blue: $4$ lowest modes, whose eigenvectors are plotted in C).
}
\end{figure*}

\section*{Local rigidity and softness at dislocations}
For many applications such as cushioning~\cite{Bilal2017}, programmed assembly~\cite{Hawkes2010}, or controlled failure~\cite{Paulose2015a}, materials need to be designed with build-in softness or rigidity. The previous section illustrates how a topological invariant can be programmed into a material to design softness at boundaries. We now proceed to show how a flux of this topological polarization can be harnessed to create soft regions in the bulk of the material. Equation~(\ref{eq:nu}) suggests that a nonzero polarization flux can be achieved by considering an inhomogeneous material.
We choose dislocations to provide this inhomogeneity.

The natural defects in three-dimensional crystals are line dislocations: displacements of unit cells along straight lines which terminate at material boundaries. In many crystalline solids, such defects control mechanical deformations and plasticity. As these lines are topologically protected, the only way to confine a line dislocation to a finite region is to form a closed \emph{dislocation loop} (Fig.~\ref{Fig4}A--B).  The topological invariant characterizing a dislocation is the Burgers vector $\bm b$, which expresses the effect of a dislocation on the surrounding lattice. If a particle at point $\bm x$ is displaced to $ \bm x+\bm u(\bm x)$ according to displacement $\bm u(\bm x)$, then the Burgers vector is $\bm b=\oint_C d\bm u$, where $C$ is any closed contour surrounding the dislocation line.

Line dislocations come in two primary types called \emph{edge} and \emph{screw} dislocations, which are distinguished by the orientation of the Burgers vector $\bm b$ relative to the dislocation line direction $\hat{\bm\ell}$ (see Fig.~\ref{Fig4}A). For edge dislocations, these vectors are orthogonal: the displacement $\bm u$ pushes unit cells apart in order to insert a half-plane of unit cells that extends from the dislocation line in the direction $\hat{\bm\ell}\times\bm b$. In this way, edge dislocations are three-dimensional generalizations of two-dimensional point dislocations. 
By contrast, for screw dislocations, the $\bm b$ and $\hat{\bm\ell}$ vectors are parallel: the displacement $\bm u$ pushes neighboring cells apart along the loop. In this way, screw dislocations give rise to an inherently three-dimensional spiral structure. Along a dislocation loop, the Burgers vector $\bm b$ is constant, but the line direction $\hat{\bm\ell}$ changes. As Fig.~\ref{Fig4}A shows, a dislocation loop can contain both edges and screws.

The interplay between line properties $(\bm{\hat{\ell}},\bm{b})$ and the topological polarization $\bm P$ of the lattice leads to a rigidity charge that can be expressed in a simple formula (derived in the SI) for the charge line density $\rho^L$ localized at the dislocation line: 
\begin{equation}
\rho^L = \frac{1}{V_{cell}} \bm P\cdot(\bm b\times\bm{\hat\ell}),
\label{eq:rhol}
\end{equation}
where  $V_{cell}$ is the unit cell volume of the uniform lattice.
Equivalently, we could write $\rho^L = V_{cell}^{-1}\det (\bm P,\bm b,\bm{\hat\ell})$. Consequently, for any configuration in which these three vectors are not linearly independent, the charge density is zero. In particular, screw dislocations do not carry charge---their Burgers vector points along the line direction. For edge dislocations, the charge is negative (positive) if $\bm P$ points along (against) the half-plane direction $\hat{\bm\ell}\times\bm b$: intuitively, these rows carry extra polarization out or in.
Note that for any dislocation loop, the net charge $\oint_L \rho^L$ is zero. 
However, dislocations loops do separate rigidity charges in space, resulting in a topological charge dipole, as shown in Fig.~\ref{Fig4}A. 
This dipole is oriented within the $\bm{b}\,$-$\bm{P}$ plane and can be quantified by a dipole moment, which captures the amount of charge separated and the distance of separation. 
The dipole's $\bm{b}$ ($\bm{P}$) component is the area of the dislocation loop after it is projected onto the plane formed by the two vectors $\bm{b}$ ($\bm{P}$) and $\bm{b}\times\bm{P}$ (see SI for derivation). For the loop composed of edge and screw dislocations
shown in Fig.~\ref{Fig4}A, all of the charges are localized along the edge dislocations and the dipole moment lies along $\bm b$. 

A positive charge density corresponds to soft modes localized along part of the dislocation loop. We investigate this localized softness within a polarized lattice using the configuration shown schematically in Fig.~\ref{Fig4}A and plotted for a small sample in Fig.~\ref{Fig4}B. In Fig.~\ref{Fig4}C, we demonstrate that for the softest (i.e., lowest-frequency non-translational) modes of this lattice, the unit cells with the largest displacements are localized along the near side of the loop, in agreement with Eq.~(\ref{eq:rhol}).
This can be contrasted with the lowest-energy modes of a sample without a dislocation: in Fig.~\ref{Fig4}D, we show that the dislocated lattice has soft modes at lower frequencies.
Whereas the lowest modes in the non-dislocated sample are the largest-wavelength acoustic phonons that fit within the periodic box, the lowest modes of the dislocated lattice are a combination of these acoustic phonons
and the many localized modes such as the ones shown in Fig.~\ref{Fig4}C.

\section*{Conclusions and outlook}
We demonstrated that dislocation loops in polarized three-dimensional lattices
can localize soft modes at one part of the loop. Similarly, the opposite part of the loop localizes states of self-stress. Although the total topological charge along the loop is zero, the loop acts as a topological charge dipole. The loop can be used to localize topologically protected soft modes in one region of the material, inside the material bulk. Furthermore, because the lattice we have designed is gapped, these localized soft modes do not compete with the many extended, periodic soft modes characteristic of Weyl loops.

Let us conclude by discussing applications for both three-dimensional gapped lattices and for localized softness. Three-dimensional topological materials may be useful in the design of devices for cushioning, including safety devices such as helmets, in which a hard side necessary to resist stress exists in combination with a soft side necessary to cushion the body~\cite{Bilal2017}.
Localized softness can be used to pre-program large displacements and to isolate the rest of the material from strain. In active mechanical metamaterials~\cite{Khanikaev2015,Yang2015, Souslov2017}, for example in self-folding origami~\cite{Hawkes2010, Silverberg2014, Chen2016}, these mechanisms could be actuated using motors to strain the material in a well-controlled way~\cite{Paulose2015}. Furthermore, localized states of self-stress can be used to control material failure via either buckling or fracture, which then isolates the rest of the material from failure even under a large external load.

\section*{Acknowledgments}
We thank Paul Baireuther and Bryan G.~Chen for fruitful discussions and T.~C.~Lubensky for a critical reading of the manuscript. 
We gratefully acknowledge funding from FOM, NWO, and Delta Institute for theoretical physics.

\clearpage

\makeatletter
\def\@seccntformat#1{%
  \expandafter\ifx\csname c@#1\endcsname\c@section\else
  \csname the#1\endcsname\quad
  \fi}
\makeatother

\setcounter{equation}{0}
\setcounter{figure}{0}

\section{Supporting Information for ``Localizing softness and stress along loops in three-dimensional topological metamaterials''}

\vspace{2em}

\renewcommand{\theequation}{S\arabic{equation}}
\renewcommand{\figurename}{{\bf Fig. }}
\renewcommand{\thefigure}{{\bf S\arabic{figure}}}

\section{Generalized stacked kagome lattice}

The generalized stacked kagome lattice is built by repeating a given unit cell along the following lattice vectors
\begin{equation}\label{eq:PrimVectors}
\begin{array}{ccc}
\bm a_1=a(1,0,0)
&\bm a_2=a(1/2,\sqrt3/2,0)
&\bm a_3=a(0,0,1)
\end{array},
\end{equation}
where $a$ defines the lattice spacing. In the undeformed kagome lattice the equilibrium positions of the particles in the unit cell is given by the triangular prism
\begin{equation}
\begin{array}{ccc}
\bm r(1)=-\tfrac{1}{4}\bm a_3
&\bm r(2)=\tfrac{1}{2}\bm a_1-\tfrac{1}{4}\bm a_3
&\bm r(3)=\tfrac{1}{2}\bm a_2-\tfrac{1}{4}\bm a_3
\vspace{4pt}\\
\bm r(4)=+\tfrac{1}{4}\bm a_3
&\bm r(5)=\tfrac{1}{2}\bm a_1+\tfrac{1}{4}\bm a_3
&\bm r(6)=\tfrac{1}{2}\bm a_2+\tfrac{1}{4}\bm a_3,
\end{array}
\end{equation}
where particles $s=1,2,3$ ($s=4,5,6$) constitute the bottom (top) triangle of the prism.

Within the unit cell we connect the particle pairs (2,3), (3,1), (1,2), (5,6), (6,4), (4,5), (1,4), (2,5), (3,6) by springs. Horizontally between unit cells we connect pairs (3,2), (1,3), (2,1), (6,5), (4,6), (5,4) according to the nearest neighbor principle. Finally we connect (4,1), (5,2) and (6,3) between the unit cells in the positive vertical direction.

We let a deformation act on the unit cell by rotating the triangles around three orthogonal axes attached to the triangles. The first two axes are coplanar to the triangle, with the first one bisecting one of its angles. The third axis is perpendicular to the triangle. The three axes meet in the center of the triangle. We then define our deformation by rotating the top triangle around the three axis by angles $\phi$, $\theta$, and $\psi$ respectively, while rotating the bottom triangle by $-\phi$, $-\theta$, and $-\psi$. Note that by taking our rotation axes relative to the triangles, our rotation group is commutative and we do not need to worry about the order of rotations. Explicitly, the equilibrium positions of the particles in the generalized stacked kagome can be given as
\begin{widetext}
\begin{align}
\bm r(1)&=\frac{a}{2\sqrt{3}}
\begin{pmatrix}
s_\psi c_\theta-\sqrt{3}c_\psi c_\phi-c_\psi s_\phi s_\theta
\\-c_\psi c_\theta-\sqrt{3} s_\psi c_\phi- s_\psi s_\phi s_\theta
\\- c_\phi s_\theta+ \sqrt{3}s_\phi+ \sqrt{3}
\end{pmatrix}+\frac{ar}{2}\begin{pmatrix}c_\alpha\\ s_\alpha\\0\end{pmatrix}\nonumber
\\\bm r(2)&=\frac{a}{2\sqrt{3}}
\begin{pmatrix}
s_\psi c_\theta+\sqrt{3}c_\psi c_\phi-c_\psi s_\phi s_\theta 
\\-c_\psi c_\theta+ \sqrt{3}s_\psi c_\phi- s_\phi s_\phi s_\theta
\\- c_\phi s_\theta- \sqrt{3}s_\phi+ \sqrt{3}
\end{pmatrix}+\frac{ar}{2}\begin{pmatrix}c_\alpha\\ s_\alpha\\0\end{pmatrix}\nonumber
\\\bm r(3)&=\hspace{5pt}\frac{a}{\sqrt{3}}
\begin{pmatrix}
- s_\psi c_\theta+c_\psi s_\phi s_\theta
\\c_\psi c_\theta+ s_\psi s_\phi s_\theta
\\ c_\phi s_\theta + \sqrt{3}/2
\end{pmatrix}+\frac{ar}{2}\begin{pmatrix}c_\alpha\\ s_\alpha\\0\end{pmatrix}\nonumber
\\\bm r(4)&=\frac{a}{2\sqrt{3}}
\begin{pmatrix}
- s_\psi c_\theta-\sqrt{3}c_\psi c_\phi-c_\psi s_\phi s_\theta
\\-c_\psi c_\theta+\sqrt{3} s_\psi c_\phi+ s_\psi s_\phi s_\theta
\\ c_\phi s_\theta- \sqrt{3}s_\phi- \sqrt{3}
\end{pmatrix}-\frac{ar}{2}\begin{pmatrix}c_\alpha\\ s_\alpha\\0\end{pmatrix}\nonumber
\\\bm r(5)&=\frac{a}{2\sqrt{3}}
\begin{pmatrix}
- s_\psi c_\theta+\sqrt{3}c_\psi c_\phi-c_\psi s_\phi s_\theta 
\\-c_\psi c_\theta- \sqrt{3}s_\psi c_\phi+ s_\phi s_\phi s_\theta
\\ c_\phi s_\theta+ \sqrt{3}s_\phi- \sqrt{3}
\end{pmatrix}-\frac{ar}{2}\begin{pmatrix}c_\alpha\\ s_\alpha\\0\end{pmatrix}\nonumber
\\\bm r(6)&=\hspace{5pt}\frac{a}{\sqrt{3}}
\begin{pmatrix}
 s_\psi c_\theta+c_\psi s_\phi s_\theta
\\c_\psi c_\theta- s_\psi s_\phi s_\theta
\\- c_\phi s_\theta - \sqrt{3}/2
\end{pmatrix}-\frac{ar}{2}\begin{pmatrix}c_\alpha\\ s_\alpha\\0\end{pmatrix},
\end{align}
\end{widetext}
where $s_\cdot=\sin(\cdot)$ and $c_\cdot=\cos(\cdot)$.

\clearpage
\begin{widetext}
Following the above order of bond enumeration, the compatibility matrix takes the form:
\begin{equation}
\label{eq:RigMat}
R^\dagger(\bm k)=\begin{pmatrix}
\begin{array}{c}
\bm{\hat b}^\mathsf{T}_1\\\bm{\hat b}^\mathsf{T}_2\\\bm{\hat b}^\mathsf{T}_3\\
\bm{\hat b}^\mathsf{T}_4\\\bm{\hat b}^\mathsf{T}_5\\\bm{\hat b}^\mathsf{T}_6\\
\bm{\hat b}^\mathsf{T}_7\\\bm{\hat b}^\mathsf{T}_8\\\bm{\hat b}^\mathsf{T}_9\\
\bm{\hat b}^\mathsf{T}_{10}\\\bm{\hat b}^\mathsf{T}_{11}\\\bm{\hat b}^\mathsf{T}_{12}\\
\bm{\hat b}^\mathsf{T}_{13}\\\bm{\hat b}^\mathsf{T}_{14}\\\bm{\hat b}^\mathsf{T}_{15}\\
\bm{\hat b}^\mathsf{T}_{16}\\\bm{\hat b}^\mathsf{T}_{17}\\\bm{\hat b}^\mathsf{T}_{18}
\end{array}&
\begin{array}{ccc|ccc}
0&-1&1&\phantom{\hat b^\mathsf{T}_1}\\
1&0&-1&\phantom{\hat b^\mathsf{T}_1}&O\\
-1&1&0&\phantom{\hat b^\mathsf{T}_1}\\\hline
\phantom{\hat b^\mathsf{T}_1}&&&0&-1&1\\
\phantom{\hat b^\mathsf{T}_1}&O&&1&0&-1\\
\phantom{\hat b^\mathsf{T}_1}&&&-1&1&0\\\hline
\phantom{\hat b^\mathsf{T}_1}&&&&&\\
\phantom{\hat b^\mathsf{T}_1}&-I&&&I&\\
\phantom{\hat b^\mathsf{T}_1}&&&&&\\\hline
0&e^{-i k_1}&-1&\phantom{\hat b^\mathsf{T}_1}\\
-1&0&e^{-i k_2}&\phantom{\hat b^\mathsf{T}_1}&O\\
e^{-i (k_1-k_2)}&-1&0&\phantom{\hat b^\mathsf{T}_1}\\\hline
\phantom{\hat b^\mathsf{T}_1}&&&0&e^{-i k_1}&-1\\
\phantom{\hat b^\mathsf{T}_1}&O&&-1&0&e^{-i k_2}\\
\phantom{\hat b^\mathsf{T}_1}&&&e^{-i(k_1-k_2)}&-1&0\\\hline
\phantom{\hat b^\mathsf{T}_1}&&&&&\\
\phantom{\hat b^\mathsf{T}_1}&e^{-i k_3}I&&&-I&\\
\phantom{\hat b^\mathsf{T}_1}&&&&&\\
\end{array}
\end{pmatrix}
\end{equation}
\end{widetext}
where the initial spring direction $\bm b_i$ is defined as the normalization of $\bm r(s')-\bm r(s)$, where $(s,s')$ is the $i$-th spring. In this notation, each number in the matrix grid represents a multiple of the triple on the far left in the same row. This turns the matrix into an $18\times18$ square matrix. As the initial bond directions depend smoothly on the initial particle positions, each element of $Q$ depends smoothly on the deformation parameters $\phi$, $\theta$, and $\psi$.

\section{Soft directions}

Weyl lines are formed by collections of points satisfying $\det R(\bm k)=0$. This equation is often quite hard to solve. However, with observational evidence suggesting that the Weyl loops generally connect to the origin, we can formally expand the determinant around $k\equiv||\bm k||=0$ as
\begin{equation}
\det R(\bm k)
=\sum_{\bm{n}\in \mathbb{Z}^3}c_{\bm n}e^{-i\bm n\cdot\bm k}
=i k^3\mathcal P_3(\hat{\bm k})+k^4\mathcal P_4(\hat{\bm k})+\mathcal O(k^5),
\end{equation}
for finitely many nonzero coefficients $c_{\bm n} \in \mathbb{R}$, where $\mathcal P_m$ is an $m$-th order homogeneous polynomial with real coefficients in three variables over the unit sphere $||\hat{\bm k}||=1$. The vanishing constant, linear, and quadratic polynomials reflect the triple multiplicity of the zero at $\bm k=0$.

A Weyl line can only attach to the origin at a direction $\hat{\bm k}$ where both $\mathcal P_3(\hat{\bm k})$ and $\mathcal P_4(\hat{\bm k})$ vanish. 
Suppose, for instance, that $\mathcal P_4(\hat{\bm k}) \ne 0$ and note that the error term is dominated by $M k^5$ for all $k<\delta$, where $M$ and $\delta$ are fixed positive numbers.
As a consequence, the relation
\begin{equation}
|\det R(\bm k)|> k^4\left(|\mathcal P_4(\hat{\bm k})|-M k\right)
\end{equation}
holds for all $k<\delta$. In particular, this means $|\det R(\bm k)|>0$ for all $k<\min\left(\delta,\frac{1}{M}|\mathcal P_4(\hat{\bm k})|\right)$. We call these common zeroes of $\mathcal P_3(\hat{\bm k})$ and $\mathcal P_4(\hat{\bm k})$ over the unit sphere the soft directions of $\det R$. 

As any Weyl loop connects to the origin along two directions, we find that a configuration with $2n$ soft directions can accommodate no more than $n$ pairs of Weyl loops. 
In the main text we show the $\theta=0$ slice of soft-mode phase space, containing a region where the lattice does not exhibit any soft directions. This absence of soft directions is rare. Sampling of the $0<\phi,\theta,\psi<\pi/2$ phase space with step size $\pi/60$ we obtain the following data for the prevalence of different numbers of soft directions:
\begin{table}[ht!]\centering
\begin{tabular}{lr|r|r|r|r|r|r}
Number of soft directions:&0&2&4&6&8&10&12\\\hline
Percentage of phase space:\phantom{$\Big|$}&$<$0.1&20.0&53.5&17.3&6.9&2.2&$<$0.1
\end{tabular}
\end{table}

\section{Rigidity charge bound to line dislocations}

The dislocation acts on the polarization and the unit cell volume through the lattice vectors $\bm a_i$. Letting the displacement function $\bm u$ act on both the initial and terminal point of the vector $\bm a_i$, we obtain the lattice vector field in the dislocated lattice:
\begin{equation}
\tilde{\bm a}_i(\bm x)
=[\bm x+\bm a_i+\bm u(\bm x+\bm a_i)]-[\bm x+\bm u(\bm x)]
=\bm a_i+(\bm a_i\cdot\bm\nabla)\bm u(\bm x),
\end{equation}
where we assume that higher order differentials of $\bm u$ are negligible.
The polarization changes as
\begin{equation}
\tilde{\bm P}(\bm x)
=\sum_i m_i\tilde{\bm a}_i
=\bm P+(\bm P\cdot\bm\nabla)\bm u(\bm x).
\end{equation}
The unit cell volume changes as
\begin{equation}
\tilde{V}_{cell}(\bm x)
=\det( \tilde{\bm a}_1,\tilde{\bm a}_2,\tilde{\bm a}_3)
=V_{cell}(1+\bm\nabla\bm u(\bm x)).
\end{equation}
We start from Eq.~(1) in the main text. In a canonical basis with coordinates $(x^1,x^2,x^3)$, a surface element is given by $dx^\ell\wedge dx^k=\epsilon^{i\ell k}dS_i$, where $d\bm S$ is positively oriented. In this notation, the change in rigidity charge in the region $S$ from the $\nu^S=0$ periodic case is given by
\begin{align}
\tilde{\nu}^S-\nu^S
&=\int_{\partial S}(\tilde{V}_{cell}^{-1}\,\tilde{\bm P}-V_{cell}^{-1}\,\bm P)\cdot(-d\bm S)\nonumber\\&
=V_{cell}^{-1}\int_{\partial S} P_i\partial_ju_j-P_j\partial_ju_i\,dS_i\nonumber\\&
=V_{cell}^{-1}\int_{\partial S}P_i \epsilon_{ijk}\partial_\ell u_j\,\epsilon^{m\ell k}dS_m\nonumber\\&
=V_{cell}^{-1}\int_{\partial S}\big(\bm P\times d\bm u\big)_k\wedge dx^k. \label{eq:dislcharge}
\end{align}

Consider a straight dislocation line segment $L$ in direction $\hat{\bm\ell}$ and let $S$ be the cylinder with axis $L$ and radius $\mathcal R$. Without loss of generality, we choose canonical coordinates such that $L=\{t\hat{\bm\ell}: 0<t<T\}$ with $\hat{\bm\ell}$ pointing along the third basis vector. Furthermore, we take $T\hat{\bm\ell}$ to be an integer combination of lattice vectors $\bm a_i$. If the displacement function respects the periodicity of the lattice, the top ($t=T$) and bottom ($t=0$) surfaces of $S$ will cancel by orientation. It remains to calculate Eq.~(\ref{eq:dislcharge}) along the vertical boundary of $S$. First consider the $k=1$ term:
\begin{equation}
\int_{\partial S}\big(\bm P\times d\bm u\big)_1\wedge dx^1=\int_0^{2\pi}\left(\bm P\times\textstyle\int_{L_\theta}d\bm u\right)_1\,d\mathcal R\cos\theta=0,
\end{equation}
where $\partial S=\bigcup_\theta L_\theta$ is a decomposition of the cylindrical boundary in vertical lines at different angle $\theta$. The last equality holds since $\int_{L_\theta}d\bm u$ is independent of $\theta$ as can be shown by contour integration
\begin{equation}
\int_{L_\theta}d\bm u-\int_{L_{\theta'}}d\bm u=\int_\theta^{\theta'}d\bm u\big|^{t=T}-d\bm u\big|^{t=0}=0.
\end{equation}
The $k=2$ term vanishes similarly. Hence we find that the rigidity charge is given entirely by the term corresponding to the line direction. Let $\partial S=\bigcup_t C_t$ be the decomposition of the cylindrical boundary in horizontal circles at different height $t$. Recalling the definition of the Burgers vector, we then obtain
\begin{equation}
\tilde{\nu}^S
=V_{cell}^{-1}\int_0^T\left(\bm P\times\textstyle\int_{C_t}d\bm u\right)_3\, dt=\displaystyle\frac{1}{V_{cell}}(\bm P\times\bm b)\cdot T\hat{\bm\ell}.
\end{equation}
The rigidity charge density along the line can then be given by $\rho_L\equiv\nu_L/T=V_{cell}^{-1}\det(\bm P,\bm b,\hat{\bm\ell})$, where the three vectors $\bm P$, $\bm b$ and $\hat{\bm\ell}$ form the columns of a $3\times3$ matrix.

\section{Dislocation loop dipole moment}

Let $\bm r(s)$ be an arc length parametrization of the loop $L$. The total charge along the loop is given by
\begin{equation}
\nu^L=\oint_L\rho^L(\bm r)ds\propto\oint_L \bm{r}'(s)ds=0.
\end{equation}
As the loop separates charges of different sign, it is sensible to consider the associated dipole moment
\begin{equation}
\bm d^L\equiv\oint_L \bm r\rho^L(\bm r)ds=\oint_L \bm r(s)\frac{(\bm P\times\bm b)\cdot \bm r'(s)}{V_{cell}}ds.
\end{equation}
Without loss of generality we can choose our basis $(\hat{\bm e}_1,\hat{\bm e}_2,\hat{\bm e}_3)$ such that $\bm P\times\bm b$ points in the direction of $\hat{\bm e}_3$. We then obtain:
\begin{equation}
\bm{d}^L=\frac{||\bm P\times\bm b||}{V_{cell}}\oint_L \bm{r} \, dr_3
=\frac{||\bm P\times\bm b||}{V_{cell}}(\mathcal A_{13},\mathcal A_{23}, 0),
\end{equation}
where $\mathcal A_{ij}$ is the signed area of the projection of the dislocation loop on the plane generated by the basis $(\hat{\bm e}_i,\hat{\bm e}_j)$.


\clearpage

\begin{thebibliography}{10}

\bibitem{Bertoldi2010}
Bertoldi K, Reis PM, Willshaw S, Mullin T (2010) Negative poisson's ratio
  behavior induced by an elastic instability.
\newblock {\em Advanced Materials} 22(3):361--366.

\bibitem{Sun2012}
Sun K, Souslov A, Mao X, Lubensky TC (2012) Surface phonons, elastic response,
  and conformal invariance in twisted kagome lattices.
\newblock {\em Proceedings of the National Academy of Sciences}
  109(31):12369--12374.

\bibitem{Kane2014}
Kane CL, Lubensky TC (2013) {Topological boundary modes in isostatic lattices}.
\newblock {\em Nature Physics} 10(1):39--45.

\bibitem{Chen2014}
Chen BGg, Upadhyaya N, Vitelli V (2014) {Nonlinear conduction via solitons in a
  topological mechanical insulator}.
\newblock {\em Proceedings of the National Academy of Sciences}
  111(36):13004--13009.

\bibitem{Lubensky2015}
Lubensky TC, Kane CL, Mao X, Souslov A, Sun K (2015) Phonons and elasticity in
  critically coordinated lattices.
\newblock {\em Reports on Progress in Physics} 78(7):073901.

\bibitem{Meeussen2017}
Meeussen AS, Paulose J, Vitelli V (2016) Geared topological metamaterials with
  tunable mechanical stability.
\newblock {\em Phys. Rev. X} 6(4):041029.

\bibitem{Rocks2017}
Rocks JW, et~al. (2017) Designing allostery-inspired response in mechanical
  networks.
\newblock {\em Proceedings of the National Academy of Sciences}
  114(10):2520--2525.

\bibitem{Yan2017}
Yan L, Ravasio R, Brito C, Wyart M (2017) Architecture and coevolution of
  allosteric materials.
\newblock {\em Proceedings of the National Academy of Sciences}
  114(10):2526--2531.

\bibitem{Coulais2017}
Coulais C, Sounas D, Al{\`u} A (2017) Static non-reciprocity in mechanical
  metamaterials.
\newblock {\em Nature} 542(7642):461--464.

\bibitem{Maxwell1864}
Maxwell JC (1864) On the calculation of the equilibrium and stiffness of
  frames.
\newblock {\em Philosophical Magazine} 27(182):294--299.

\bibitem{Calladine1978}
Calladine C (1978) Buckminster fuller's “tensegrity” structures and clerk
  maxwell's rules for the construction of stiff frames.
\newblock {\em International Journal of Solids and Structures} 14(2):161 --
  172.

\bibitem{Guest2003}
Guest S, Hutchinson J (2003) On the determinacy of repetitive structures.
\newblock {\em Journal of the Mechanics and Physics of Solids} 51(3):383 --
  391.

\bibitem{Kapko2009}
Kapko V, Treacy MMJ, Thorpe MF, Guest SD (2009) On the collapse of locally
  isostatic networks.
\newblock {\em Proceedings of the Royal Society of London A: Mathematical,
  Physical and Engineering Sciences} 465(2111):3517--3530.

\bibitem{Rocklin2016}
Rocklin DZ, Chen BGg, Falk M, Vitelli V, Lubensky TC (2016) {Mechanical Weyl
  Modes in Topological Maxwell Lattices}.
\newblock {\em Physical Review Letters} 116(13):135503.

\bibitem{Stenull2016}
Stenull O, Kane C, Lubensky T (2016) Topological phonons and weyl lines in
  three dimensions.
\newblock {\em Physical Review Letters} 117(6):068001.

\bibitem{Florijn2014}
Florijn B, Coulais C, van Hecke M (2014) Programmable mechanical metamaterials.
\newblock {\em Phys. Rev. Lett.} 113(17):175503.

\bibitem{Lakes1987}
Lakes R (1987) Foam structures with a negative poisson{\textquoteright}s ratio.
\newblock {\em Science} 235(4792):1038--1040.

\bibitem{Rocklin2016a}
Rocklin D (2016) Directional mechanical response in the bulk of topological
  metamaterials.
\newblock {\em arXiv preprint arXiv:1612.00084}.

\bibitem{Sussman2016}
Sussman DM, Stenull O, Lubensky TC (2016) Topological boundary modes in jammed
  matter.
\newblock {\em Soft Matter} 12(28):6079--6087.

\bibitem{Nash2015}
Nash LM, et~al. (2015) {Topological mechanics of gyroscopic metamaterials.}
\newblock {\em Proc. Natl. Acad. Sci. USA} 112(47):14495--500.

\bibitem{Susstrunk2015}
Susstrunk R, Huber SD (2015) {Observation of phononic helical edge states in a
  mechanical topological insulator}.
\newblock {\em Science} 349(6243):47--50.

\bibitem{Kariyado2015}
Kariyado T, Hatsugai Y (2015) Manipulation of dirac cones in mechanical
  graphene.
\newblock {\em Scientific reports} 5:18107.

\bibitem{Mousavi2015}
Mousavi SH, Khanikaev AB, Wang Z (2015) {Topologically protected elastic waves
  in phononic metamaterials.}
\newblock {\em Nature communications} 6:8682.

\bibitem{Khanikaev2015}
Khanikaev AB, Fleury R, Mousavi SH, Al{\`{u}} A (2015) {Topologically robust
  sound propagation in an angular-momentum-biased graphene-like resonator
  lattice}.
\newblock {\em Nature Communications} 6(May):8260.

\bibitem{Yang2015}
Yang Z, et~al. (2015) {Topological Acoustics}.
\newblock {\em Physical Review Letters} 114(11):114301.

\bibitem{Huber2016}
Huber SD (2016) Topological mechanics.
\newblock {\em Nature Physics} 12(7):621--623.

\bibitem{Rechtsman2013}
Rechtsman MC, et~al. (2013) {Photonic Floquet topological insulators.}
\newblock {\em Nature} 496(7444):196--200.

\bibitem{Lu2014}
Lu L, Joannopoulos JD, Solja{\v{c}}i{\'c} M (2014) Topological photonics.
\newblock {\em Nature Photonics} 8(11):821--829.

\bibitem{Wang2015}
Wang P, Lu L, Bertoldi K (2015) {Topological Phononic Crystals with One-Way
  Elastic Edge Waves.}
\newblock {\em Physical review letters} 115(10):104302.

\bibitem{Hasan2010}
Hasan MZ, Kane CL (2010) {Colloquium: Topological insulators}.
\newblock {\em Reviews of Modern Physics} 82(4):3045--3067.

\bibitem{Bernevig2013}
Bernevig BA, Hughes TL (2013) {\em {Topological Insulators and Topological
  Superconductors}}.

\bibitem{Rocklin2017}
Rocklin DZ, Zhou S, Sun K, Mao X (2017) Transformable topological mechanical
  metamaterials.
\newblock {\em Nature communications} 8:14201.

\bibitem{Paulose2015}
Paulose J, Chen BGg, Vitelli V (2015) {Topological modes bound to dislocations
  in mechanical metamaterials}.
\newblock {\em Nature Physics} 11:153--156.

\bibitem{Paulose2015a}
Paulose J, Meeussen AS, Vitelli V (2015) {Selective buckling via states of
  self-stress in topological metamaterials}.
\newblock {\em Proceedings of the National Academy of Sciences}
  112(25):7639--7644.

\bibitem{Bilal2017}
Bilal OR, Susstrunk R, Daraio C, Huber SD (2017) Intrinsically polar elastic
  metamaterials.
\newblock {\em Advanced Materials} p. 1700540.

\bibitem{Volovik2007}
Volovik GE (2007) {\em Quantum Phase Transitions from Topology in Momentum
  Space}, eds.{} Unruh WG, Sch{\"u}tzhold R.
\newblock (Springer Berlin Heidelberg, Berlin, Heidelberg), pp. 31--73.

\bibitem{Burkov2011}
Burkov AA, Hook MD, Balents L (2011) Topological nodal semimetals.
\newblock {\em Phys. Rev. B} 84(23):235126.

\bibitem{Lu2013}
Lu L, Fu L, Joannopoulos JD, Solja{\v{c}}i{\'c} M (2013) Weyl points and line
  nodes in gyroid photonic crystals.
\newblock {\em Nature photonics} 7(4):294--299.

\bibitem{Xu2015}
Xu SY, et~al. (2015) Discovery of a weyl fermion semimetal and topological
  fermi arcs.
\newblock {\em Science} 349(6248):613--617.

\bibitem{Lu2015}
Lu L, et~al. (2015) Experimental observation of weyl points.
\newblock {\em Science} 349(6248):622--624.

\bibitem{Hawkes2010}
Hawkes E, et~al. (2010) Programmable matter by folding.
\newblock {\em Proceedings of the National Academy of Sciences}
  107(28):12441--12445.

\bibitem{Souslov2017}
Souslov A, van Zuiden BC, Bartolo D, Vitelli V (2017) Topological sound in
  active-liquid metamaterials.
\newblock {\em Nature Physics}.

\bibitem{Silverberg2014}
Silverberg JL, et~al. (2014) Using origami design principles to fold
  reprogrammable mechanical metamaterials.
\newblock {\em Science} 345(6197):647--650.

\bibitem{Chen2016}
Chen BGG, et~al. (2016) {Topological Mechanics of Origami and Kirigami}.
\newblock {\em Physical Review Letters} 116(13):135501.

\end{thebibliography}
\end{document}